\documentstyle[preprint,aps]{revtex}  

\tighten       

\begin{document}
\draft 

\title{Current-voltage characteristic and  stability
in resonant-tunneling n-doped semiconductor 
superlattices}
\author{A. Wacker\cite{byline}, M. Moscoso, M. Kindelan, and L. L. Bonilla}
\address{Escuela Polit\'ecnica Superior,  Universidad Carlos III de Madrid,
Butarque 15, 28911 Legan{\'e}s, Spain}
\date{\today}
\maketitle
\begin{abstract}
We review the occurrence
of electric-field domains in doped
superlattices within a discrete drift model.
A complete analysis of the construction and stability of stationary
field profiles having two domains is carried out. As a consequence, we
can provide a simple analytical estimation  for the doping density
above which stable stable domains occur. This bound  may
be useful for the design of superlattices exhibiting
self-sustained current oscillations.
Furthermore we  explain why stable domains occur in superlattices
in contrast to the usual Gunn diode.
\end{abstract}

\section{Introduction}
Electrical transport in semiconductor superlattices (SL) has attracted much
properties related to the artificial band structure.
One of these features is the occurrence of stationary electric field domains
which have already been reported in Ref. \cite{ESA74}.
Due to advanced growing facilities and experimental techniques  the
complicated structure of the current-voltage characteristics which exhibits
several branches, roughly equal to the number of quantum wells,
could be resolved during the last decade 
\cite{KAW86,CHO87,HEL90,GRA91,KWO94,KAS94}.
In these experiments it was demonstrated that the different branches are
connected to the formation of two domains with different values of the
electric field inside the sample.
Depending on the conditions, stable stationary field domains and 
traveling domain boundaries may occur. In the latter case, the dynamics
of the electric field domains gives rise to time-dependent oscillations
of the current.\cite{MER95,KWO95a,GRA95a,KAS95}

 Early approaches towards a modelling of these phenomena
have been performed within a continuum model including strong
diffusion \cite{IGN85} and with the help of an equivalent circuit \cite{LAI91}.
During the last years it has been shown theoretically that the observed
phenomena can be fairly well reproduced by models which 
essentially combine the discrete Poisson-equation and rate 
equations for the carrier densities 
in the different quantum wells \cite{PRE94,BON94,MIL94}. Also the 
time-dependent current oscillations could be recovered in these models 
\cite{BON94,BON95a,WAC95}. A prediction of spatio-temporal chaos in 
resonant-tunneling superlattices under dc+ac voltage bias has been made on the
basis of the discrete drift model \cite{BUL95b}. The influence of growth-related
imperfections on the SL behavior has been studied in \cite{WAC95b}. 
 Some of these phenomena may also be 
described by discrete models with Monte Carlo dynamics incorporating single-electron tunneling 
effects (wich are important for slim superlattices and give rise to additional 
oscillations of the current\cite{KOR93,KOR94}).

In this paper we want to explain {\em how} these complicated phenomena
are generated by such models. This provides a deeper insight into
the basic mechanisms and helps to classify the results
of various experiments and computer simulations.
In particular we want to give an answer to the following questions:
How is it possible to understand the appearance of the
complicated structure of the current-voltage characteristic?
What are the conditions for stability and oscillations and how can
they be understood?
What is the main difference to the Gunn diode, where hardly any stable
domain states are observed?

The paper is organized as follows: The model we use is described in the second 
section. The third section shows how the complex stationary current-voltage 
characteristic changes as the doping increases.
In the fourth section we investigate the stability of the stationary states
and prove an explicit criterion for the occurrence of stable domain states.
 The last section contains our conclusions and the Appendix
is devoted to a proof that no self-sustained oscillations appear for a small
product of doping times the number of SL periods.
 When the doping density lies in the range between these two stability
boundaries, the model exhibits self-sustained oscillations.
Results concerning self-sustained oscillations of the current are presented 
in a companion paper\cite{KAS96}, where a direct comparison with  
experimental data is made.

\section{The model}
We consider a semiconductor superlattice where the lateral extension of wells
and barriers is much larger than the total length of the SL, so that
single-electron tunneling effects (see e.g. Ref. \cite{KOR93}) are negligible. 
The quantum wells (QW) are weakly coupled and the scattering times are much
shorter than the tunneling time between adjacent QWs. Thus it makes sense to 
consider the electrons to be localized within the QWs and in local equilibrium
at the lattice temperature. The current is mainly determined by the 
resonances between the different energy levels in the QWs, which we denote by
Ci, $i= 1,2,\ldots$, in order of increasing energy counted from the bottom of
the conduction band. For the biases of interest here, there are three 
important resonances C1C1, C1C2 and C1C3. If the intersubband relaxation is 
also fast with respect to the tunneling, in practice only the lowest subband 
is occupied. In this case it is a reasonable approximation 
to consider the QWs as entities characterized by average values
of the electron density $\tilde{n}_{i}$ in the $i$-th QW (in units cm$^{-2}$)
and the electric field  $\tilde{E}_{i}$ between wells $i$ and $i+1$,
with $i=1,\dots N$. For sufficiently high electric fields,
the electrons tunnel only in the  forward direction and we
can describe  the dynamical behavior of $\tilde{n}_{i}$
by the rate equation
\begin{equation}
\frac{d \tilde{n}_{i}}{d \tilde{t}}= {1\over \tilde{l}}\, \left(
\tilde{v}(\tilde{E}_{i-1})\,\tilde{n}_{i-1}-
\tilde{v}(\tilde{E}_{i})\,\tilde{n}_{i} \right)\, ,
\end{equation}
where $ \tilde{l}$ is the superlattice period and $\tilde{v}(\tilde{E})/
\tilde{l}$ is the average electron tunneling-rate for the local field. 
$\tilde{v}(\tilde{E})$ has a peak at certain values of 
the electric field connected to resonant tunneling C1C1, C1C2 and C1C3. 
In this paper we shall be concerned with phenomena occurring at fields higher
than the first resonant peak C1C2, so that we shall omit the miniband peak
C1C1 in our tunneling rate $v(E)$; see the curve plotted in Fig.~\ref{Figchars} 
by a dashed line. Phenomena at lower fields
have been studied by adding the C1C1 peak to our curve, \cite{BUL95a}.
The electric fields must fulfil the
Poisson equation averaged over one SL period
\begin{equation}
\tilde{E}_{i} - \tilde{E}_{i-1} = \frac{q}{\epsilon}\,
(\tilde{n}_{i} - \tilde{N}_{D})\, ,          \label{poisson}
\end{equation}
where $\epsilon$, $q$, and $\tilde{N}_{D}$ are the average permittivity, 
the charge of the electron, and the donor density per SL-period 
(in units cm$^{-2}$), respectively.
By differentiating Eq.~(\ref{poisson}) with respect to time
we obtain Amp\`{e}re's law for the total
current density $\tilde{J}(\tilde{t})$ \cite{BON95a}:
\begin{eqnarray}
\epsilon\,\frac{d \tilde{E}_{i}}{d \tilde{t}} +
{1\over \tilde{l}} \tilde{v}(\tilde{E}_{i})\left(q\tilde{N}_{D}+ 
\epsilon(\tilde{E}_{i} - \tilde{E}_{i-1})\right)= \tilde{J},  \label{ampere} 
\end{eqnarray}
where $i= 1,\dots, N$. Typically the experiments
are performed with a constant $dc$ voltage bias
$\tilde\Phi$ yielding
\begin{eqnarray}
 \tilde{l}\,\sum_{i=1}^{N} \tilde{E}_{i} = \tilde{\Phi}.
\label{bias}
\end{eqnarray}
Notice that there are 2N+2 unknowns: 
$$\tilde{E}_{0},\tilde{E}_{1},\ldots,
\tilde{E}_{N}, \tilde{n}_{1},\ldots,\tilde{n}_{N},\tilde{J}$$ 
and 2N+1 equations, so that we need to specify one boundary condition for 
$\tilde{E}_{0}$ plus an appropriate initial profile $\tilde{E}_{i}(0)$. 
The boundary condition for $\tilde{E}_{0}$ (the average electric field 
{\em before} the SL) can be fixed by specifying the electron density at 
the first site, $\tilde{n}_{1}$, according to (\ref{poisson}). In typical 
experiments the region before the SL has an excess of electrons due to a 
stronger n-doping there than in the SL \cite{MER95,KWO95a,GRA95a,KAS95}. 
Thus it is plausible assuming that there is an excess number of electrons 
at the first SL period measured by a dimensionless parameter $c>-1$:
\begin{equation}
\tilde{n}_1=(1+c)\tilde{N}_D\, .\label{bc}
\end{equation}
$c$ has to be quite small because it is known that a steady 
uniform-electric-field profile is observed at low laser illumination in undoped 
SL \cite{MER95,BON94,BON95a}. This observation allows us to infer the 
electron tunneling-rate $v(E)$ directly from measured current-voltage data \cite{BON94}.
Another possibility is to derive $v(E)$ from simple 
one-dimensional quantum-mechanical calculations of 
resonant tunneling, as was done by Prengel et al \cite{PRE94}. They used
a more complicated discrete model with two electron densities corresponding
to the populations of the two lower energy levels of each QW.
Their model reduces to a form of ours 
when the large separation between the time scales of phonon scattering, 
resonant tunneling and dielectric relaxation is taken into account.

For the calculations that follow, it is convenient to render the equations 
(\ref{ampere})-(\ref{bc}) dimensionless by adopting as the units of electric 
field and tunneling-rate the values at the C1C2 peak of the 
$v(E)$ curve, 
$\tilde{v}(\tilde{E})$, $\tilde{E}_{M}$ and $\tilde{v}_M$ (about $10^{5}$V/cm 
and $10^{3}$ cm/s, respectively, for the sample of Ref.\ \cite{GRA95a}).
Like in Ref.~\cite{BON95a} we set:
\begin{eqnarray}
E_i = \frac{\tilde{E}_{i}}{\tilde{E}_{M}}\, , \quad\quad n_j = 
\frac{q\tilde{n}_{i}}{\epsilon\tilde{E}_{M}}\, ,
\quad\quad I =  \frac{\tilde{J}\tilde{l}}{q\tilde{N}_{D}\tilde{v}_{M}}\, 
, \nonumber\\
v=\frac{\tilde{v}}{\tilde{v}_{M}}\, ,\quad\quad
 t = \frac{q\tilde{N}_{D}\tilde{v}_{M}\,\tilde{t}}
{\epsilon\tilde{E}_{M} \tilde{l}}\, , 
\quad\quad \phi = \frac{\tilde{\Phi}}{N\,\tilde{E}_{M}\,\tilde{l}}
\label{tilden}
\end{eqnarray}
We obtain the dimensionless equations:
\begin{equation}
\frac{dE_i(t)}{dt}=I(t)-\left(1+\frac{E_i-E_{i-1}}{\nu}\right)\, v(E_i) 
\label{evolution}
\end{equation}
for $i=1,\dots N$,
\begin{equation}
\phi=\frac{1}{N}\sum_{i=1}^N E_i(t)\, ,\label{Eqvoltage}
\end{equation}
and the boundary condition
\begin{equation}
E_0(t)=E_1(t)-c\nu . \label{dimlessbc}
\end{equation}
The dimensionless parameter $\nu$, is defined by
\begin{equation}
\nu = \frac{\tilde{N}_{D}\, q}{\epsilon\,\tilde{E}_{M}},		
\end{equation}
which yields about 0.1 for the SL used in the experiments 
\cite{GRA95a,KAS95}). 
The constant voltage condition (\ref{Eqvoltage}) determines the 
current to be
\begin{equation}
I(t)=\frac{1}{N}\sum_{i=1}^N \left(1+\frac{E_i-E_{i-1}}{\nu}\right)\, v(E_i)
\end{equation}
With the choice
(\ref{tilden}), the dimensionless tunneling rate $v(E)$ has a maximum at $E=1$ with 
$v(1)=1$. Throughout this paper we use the function $v(E)$ which is
plotted in Fig.~\ref{Figchars} by a dashed line. Besides having a maximum 
$v(1) = 1$, it has a minimum at $E_m\approx 1.667$ with $v(E_m)=v_m\approx 
0.323$. Nevertheless nearly all of the features discussed in the following are 
independent of the exact shape of the function $v(E)$. 
We only impose the restrictions that $v(E)>0$ for $E>0$ and the existence
of a minimum at $E_m>1$.

For any $\phi > 0$, we shall assume that the initial electric 
field profile is strictly positive and that the electron density is 
non-negative: $E_i(0)>0,\, n_i(0)\equiv E_i(0) - E_{i-1}(0) + \nu\geq 0,\,
\forall i$. This is reasonable unless $\phi$ is close to zero (but then the 
C1C1 peak in the tunneling rate curve should be restored) or the boundary conditions 
are unrealistic. From the equations and our assumption on the initial field 
profile, it follows that $E_i>0$ and $n_i\equiv E_i - E_{i-1} + \nu\geq 0$ 
for all positive times. The model equations have interesting properties 
concerning the monotonic behavior of the electric fields with respect
to the QW number. These  are summarized in the following Lemmas:
\bigskip

\noindent {\bf Lemma 1.} {\em If the fields of two adjacent QWs are identical,
i.e., $E_k=E_{k-1}$ holds for some $k$ with $2\le k \le N$, there is at least 
one $i$ from $1\le i \le k$ with
$E_i=E_{i-1}$ and $d(E_i-E_{i-1})/dt \neq 0$. 
For $c=0$ there is the additional
possibility that $E_0=E_1=\dots =E_k$ holds.
}
\bigskip

\noindent {\bf Lemma 2.} {\em If $c\ge 0$ and the field distribution 
is monotone increasing at $t=0$ ($E_i(0)\geq E_{i-1}(0),\, \forall i$),
it will will keep this property
for all later times $t>0$. (For $c\le 0$, the same holds for a  
decreasing field distribution).
}
\bigskip

Both lemmas can be easily proved. If $E_k=E_{k-1}$ 
holds Eq.~\ref{evolution} gives
\begin{equation}
\frac{d}{dt}(E_k-E_{k-1})=\frac{E_{k-1}-E_{k-2}}{\nu}\, v(E_{k-1}) 
\label{Eqhilf},
\end{equation}
Therefore $d(E_k-E_{k-1})/dt= 0$ implies $E_{k-1}=E_{k-2}$.
Then we may repeat the argument until we either 
find  an $i\in \{2,3,\ldots k\}$ with $E_i=E_{i-1}$ and
$d(E_i-E_{i-1})/dt \neq 0$  or we find
$E_1=E_0$ which violates the boundary condition for $c\neq 0$.
This proves lemma 1. Lemma 2 is proved by contradiction.
Assume that at $t\ge 0$ the sequence $\{E_i\}$ is increasing
but $E_k=E_{k-1}$ and $d(E_k-E_{k-1})/dt< 0$ for some $k$.
Then Eq.~(\ref{Eqhilf}) yields $E_{k-1}-E_{k-2}< 0$ in 
contradiction to the assumed increasing behavior at $t$.

In order to understand the properties of the current-voltage characteristics,
we will use repeatedly these basic lemmas in what follows.

\section{Stationary states}
In this chapter we want to explain how the complex domain
structures found experimentally and from computer simulations
\cite{GRA91,KAS94,PRE94,BON94} are generated by this simple model.

We denote the electric field profile and the current density of stationary 
states by $E^*_i$ and $I^*$, respectively. An easy way to construct the
stationary profiles is to fix $I^*$, find out the corresponding electric
field profile $\{E^*_i\}, \, i=1,\ldots,N$, calculate their voltage as a 
function of $I^*$
\begin{equation}
\phi(I^*) =\frac{1}{N}\sum_{i=1}^N E^*_i(I^*)\label{Phi-I}\, .
\end{equation}
The field profile  must fulfil the equation
\begin{equation}
E^*_{i-1}=E^*_i+\nu\left(1-\frac{I^*}{v(E^*_i)}\right)=: f(E^*_i,I^*)
\label{Eqfdef}
\end{equation}  
 which was also used in Ref.~\cite{BON94}.
The boundary condition implies 
\begin{equation}
v(E^*_1)=\frac{I^*}{c+1} \Leftrightarrow f(E^*_1,I^*)=E^*_1-c\nu
\label{Eqinitials}
\end{equation}
which has three solutions $E^*_1$ for a known fixed value of the current on 
the interval $(1+c)\, v_m<I^*<1+c$.

In order to understand the properties of the stationary profiles we will  
now investigate the behavior of the set $\{E^*_i\}$ as a function of $E^*_1$. 
At first we restrict ourselves to $c>0$ and increasing field profiles.
To construct $\{E^*_i\}$, we have to invert the function $f(E,I^*)$ for a
fixed value of $I^*$. Its derivative is:
\begin{equation}
\frac{\partial f(E,I^*)}{\partial E}=1+\nu\frac{I^*}{v(E)^2}\frac{dv(E)}{dE}
\end{equation}
With the restriction to  increasing field profiles, we can always 
obtain $E^*_i$ for $E^*_{1}\ge E_m$ because $dv/dE>0$ and $f(E_m,I^*)<E_m$ 
holds. If $E^*_1<E_m$, 
Eq.~(\ref{Eqinitials}) implies $I^*\le 1+c$ and we find that
$f(E,I^*)$ is strictly monotone increasing for all $E$ if
\begin{equation}
\frac{1}{\nu}\ge 
{\rm max}\left\{\frac{-(1+c)}{v(E)^2}\frac{dv(E)}{dE}\right\}
\label{Eqinvert}
\end{equation}
 yielding a condition which only depends on the $v(E)$ relation 
but is not dependent on $I$.
For our  function $v(E)$ this yields $\nu\le 0.195/(1+c)$.
In this case the function $f(E,I^*)$ ($I^*$ fixed) is always invertible.
Then we can find a unique field profile parametrized by the point $E^*_1$. 
Since we have three possible solutions $E^*_1$ of (\ref{Eqinitials}) for each 
given $I^*/(1+c)\in(v_m,1)$, there are three different voltages $\phi$ for each 
value of the current in this range. The function $\phi(I^*)$ is thus 
three-valued, which means that by inverting it we obtain an N- or an Z-shaped 
current-voltage characteristic as shown in Fig.~\ref{Figchars} for $\nu=0.05$ 
and $\nu=0.15$, respectively. Both types can be easily understood: When 
the doping density $\nu$ is low, Eq.~(\ref{Eqfdef}) shows that $E^*_i\approx 
E^*_1$ holds. Thus, the field profile is nearly uniform and the current-voltage 
characteristics follows the $v(E)$-curve as shown in
Fig.~\ref{Figchars} for $\nu=0.05$. This is physically obvious as there 
are few charges present inside the sample. For larger values of $\nu$ the 
values $E^*_i$ may strongly deviate from $E^*_1$ with increasing $i$, if $E^*_1$ 
is not a fixed point of $f(E,I^*)$ (which is the case for $c=0$ \cite{BON94}). 
Let us denote by $E^{(1)}(I)<E^{(2)}(I)<E^{(3)}(I)$ the three solutions of 
$v(E) = I$ for a given $I\in (v_m,1)$ which are the
fixed points of Eq.~(\ref{Eqfdef}). If $\phi$ is small 
and $c>0$, $E^*_1$ is on the first branch of $v(E)$ and the values $E^*_i$
tend to  $E^{(1)}(I^*)$. When $\phi$ is
larger and $E^*_1$ is located on the second branch of $v(E)$, the sequence 
$E^*_i$ leaves the neighborhood of $E^{(2)}(I^*)$ and then approaches $E^{(3)}(I^*)$ 
on the third branch of $v(E)$ if $\nu$ is large enough.
This is shown in Fig.~\ref{FigfofE}(a).
In this case the voltage is basicaly determined by the fixed point 
$E^{(3)}(I^*)$. Since $dE^{(3)}(I)/dI>0$, this branch of stationary solutions 
may exhibit a range of positive differential conductance leading to the Z-shape.
This effect is more pronounced for longer superlattices with 
many wells $N$ and also for larger values of $c$.

For larger doping $\nu$ the condition~(\ref{Eqinvert}) is violated
and the function $f(E,I)$ may not be invertible for some current $I$.
In this case there can be more than one possible $E_{i+1}$ following a
given $E_{i}$. Then the current-voltage characteristic can no longer
be unambiguously parametrized by the point $E_1^*$. In general $f(E,I)$ has 
3 different branches for a certain interval of $I^*$, as shown in 
Fig.~\ref{FigfofE}(b). Let us call
branch $\alpha$ that having $\partial f/\partial E>0$ for low $E$, branch 
$\beta$ has $\partial f/\partial E<0$, and branch $\gamma$ again has $\partial 
f/\partial E>0$ but for larger $E$. 

Let us explain how to construct different stationary field profiles for a 
given value of the current $I^*$. We shall assume that the profiles are 
increasing, $E^*_{i+1}\ge E^*_{i}, \,\forall i$. First of all, 
$E^*_{1}$ may be located on branch $\gamma$ of $f(E,I^*)$, and so will 
all successive fields $E^*_{i}$. This profile will have the largest possible
voltage for the same $I^*$. Secondly, $E^*_{1}$ may be located on branch 
$\beta$, which implies that all successive $E^*_{i}$ of an increasing 
field profile have to be on branch $\gamma$. The corresponding voltage is
smaller than that of the previously described branch but larger than those
stationary solution branches that we analyze next. 

If $E^*_{1}$ is on branch $\alpha$, we may have $E^*_{i}$, $i=1,\ldots,j-1$,
($j=2,\ldots,N$) on branch $\alpha$, and $E^*_{j}$ either on branch $\beta$ or on
branch $\gamma$. We obtain a different branch of stationary solutions for
each such possibility. Let us denote by $(j,\beta)$ or $(j,\gamma)$ the
solution branch having $E^*_{j}$ either on branch $\beta$ or $\gamma$ 
respectively, and $E^*_{i}$, $i=1,\ldots,j-1$ on branch $\alpha$. In Fig. 
\ref{FigfofE}(b) a solution $(j,\beta)$ is shown by a dashed line
and a solution $(j,\gamma)$ by a full line. Clearly $j=1$ corresponds to
the possibilities discussed above. Finally, we have one solution where
all field values are on branch $\alpha$ which we denote by $(N+1,\gamma)$.
In order of increasing voltages, we have
\begin{eqnarray}
\phi_{(N+1,\gamma)}(I^*)\leq \phi_{(N,\beta)}(I^*)
\leq \phi_{(N,\gamma)}(I^*)\leq
\nonumber\\
\ldots \leq \phi_{(1,\beta)}(I^*)\leq \phi_{(1,\gamma)}(I^*),\nonumber
\end{eqnarray}
corresponding to $2N+1$ different stationary solution branches with the same
current $I^*$. They can be observed in Fig.~\ref{Figchars} for $\nu=1.0$ and
$I^*=0.8$. Notice that the branches $(j+1,\gamma)$ and $(j,\beta)$ coalesce at 
a current $I^* \in (1,1+c)$ which is roughly independent of $\nu$ (see
Fig. \ref{FigfofE}(c)). The branches $(j,\beta)$ and $(j,\gamma)$ coalesce at a 
lower current $I_c$ which decreases as $\nu$ increases (see Fig. 
\ref{FigfofE}(d)). The current-voltage characteristic curve is
thus connected as shown in Fig.~\ref{Figchars} ($\nu=1.0$).

The field profile of the solution branch $(15,\gamma)$ is depicted by the
crosses in Fig.~\ref{Figfield}. One can clearly identify two regions $1\le 
i<j$ and $j<i\le N$ where the electric field $E^*_{i}$ 
is roughly constant and close to a fixed point with $v(E^*_{i})\approx I^*$.
In between there is a transition layer, the domain boundary, consisting of only
a few wells. These type of states we call domain states. 
A shift of the domain boundary by one well only changes the voltage
as long as the transition layer does not extend to one of the contacts.
As the stationary solutions resulting from a one-well shift are very similar,
the domain branches in the current--voltage characteristics look alike,
as can be seen in Fig.~\ref{Figchars} ($\nu=1.0$). The slope of the different
domain branches in Fig.~\ref{Figchars} may vary from branch to branch. 
The high field domain coresponding to far right branches has a much larger 
extension than the low field domain. Then the slope of these branches 
in Fig.~\ref{Figchars} will be closer to the slope (conductivity) of the third
branch of the curve $v(E)$. Similarly the slope of far left branches 
in Fig.~\ref{Figchars} will be closer to the slope of the first
branch of $v(E)$. The larger the difference in slope between first and 
third branches of $v(E)$ is, the larger the variation in the slope of
the branches in Fig.~\ref{Figchars} will be.

Note that the domain states are not very sensitive to the exact type
of boundary conditions if two conditions are fulfilled:
\begin{enumerate}
\item The boundary conditions must allow for the existence of a roughly
constant field distribution $E^*_{i}\approx E^{(1)}$ and  
$E^*_{i}\approx E^{(3)}$ probably after a short
contact layer of some wells. 
\item The domain boundary must be located sufficiently deep inside the sample, 
so that it does not collide with the contact layer.
\end{enumerate}

As $\nu$ decreases, the solution branches become shorter and eventually
disappear if $I_c$ becomes larger than $(1+c)$, which happens
if the inequality (\ref{Eqinvert}) holds. The stationary domain structures are
seen for narrow current intervals about $I^*=1$ 
for intermediate doping as shown in Fig.~\ref{Figchars}($\nu=0.3$).
Another complex feature can be found here for larger voltages
where extra wiggles appear. They occur if $E_1^*$ crosses the value 1, 
yielding an additional maximum in $I^*$.

Except for the fundamental question of stability, we have now understood the 
morphology of the complicated current-voltage characteristic curve shown 
in Fig.~\ref{Figchars}, its changes with doping and its relation with the
electric field profile. The very same features occur in the more complicated 
model of Prengel et al. \cite{PRE94,WAC95} as shown numerically
in Ref.~\cite{PAT95}. 

So far we have restricted ourselves to  increasing field profiles.
Therefore we could only find domain states where the high-field domain is 
located at the receiving contact and the domain boundary is an accumulation 
layer. Nevertheless, this is not the full story. For sufficiently large $\nu$ 
other solutions are also possible, even for $c>0$.
 A typical such field profile is depicted by circles in Fig.~\ref{Figfield}.
The field starts on branch $\gamma$ of the function $f(E,I^*)$ and first
increases with the QW index towards the third fixed point $E^{(3)}(I^*)$.
At a certain QW $j$ the field jumps down to either branches $\alpha$ or $\beta$,
and then decreases down towards the first fixed point $E^{(1)}(I^*)$. Of course
$\nu$ has to be large enough for these jumps down to be possible. Thus
these field profiles have a high-field domain located at the injecting contact 
and the domain boundary 
separating this domain from a low-field domain is a depletion layer. Numerical 
investigation shows that these stationary states are stable and that they can be 
reached from many initial conditions.
Lemma 2 tells us that the initial field distribution can not be monotone 
increasing in the well index (like the solutions of the connected branch 
discussed before), for otherwise the field distribution would stay  
increasing for all times. Thus, these different stationary solutions are not 
connected to the branches dicussed before (having an increasing 
field profile) but form many additional isolated closed curves or ``isolas''
\cite{IOO80,DEL82}. A typical isolated curve is shown inside the frame in 
Fig.~\ref{Figchars} for $\nu = 1.0$, which is also blown up to an enlarged 
scale for the sake of clarity.

A special situation arises if $c=0$ holds. In this case 
the branches of  increasing and decreasing field profiles  
become connected and there appears much additional degeneracy leading to an 
extremely complicated structure as observed in Ref.~\cite{BON94}.

\section{Stability of the stationary states}
Up to now we have only discussed the existence of stationary states, but
not their stability properties. First of all, several stability properties
can be established by topological arguments \cite{IOO80,WAC95a}. If several
branches overlap at a fixed voltage, each second branch has to be unstable
by general reasons. For example the middle branch (exhibiting positive
differential conductivity) of the Z-shaped characteristics in 
Fig.~\ref{Figchars}($\nu=0.15$) has to be unstable. 
The remaining branches may exhibit further bifurcations.
In order to elucidate this, we will perform a linear stability analysis for 
the states constructed in the previous section. By this method we will prove 
the following statements:
\begin{itemize}
\item[A.] For large doping $\tilde{N}_D$ exceeding approximately 
$\epsilon(\tilde{E_m} - \tilde{E_M})/(q)
\cdot \tilde{v_m}/(\tilde{v_M}-\tilde{v_m})$ we find stable
domain states.
\item[B.] For very small products $\tilde{N}_D (N-1)$ 
the almost uniform states are  also stable.
\end{itemize}
For a medium range of doping in between these two limits 
we find self generated oscillations as reported in Ref.~\cite{GRA95a,KAS95},
and further discussed in a companion paper \cite{KAS96}.

In order to perform the linear stability analysis, we substitute
\begin{eqnarray}
E_i(t)&=& E^*_i + e^{\lambda t}\,\hat{e}_i\\
I(t) &=& I^*+ e^{\lambda t}\,\hat{j}
\end{eqnarray}
in (\ref{evolution}) and (\ref{dimlessbc}), thereby obtaining
\begin{eqnarray}
\lambda \hat{e}_i = \hat{j} -\frac{I^* v'(E_i^*)}{v(E_i^*)} \hat{e}_i -
\frac{v(E_i^*)}{\nu} (\hat{e}_i-\hat{e}_{i-1}) , \label{Eqlinstab}\\
\hat{e}_0=\hat{e}_1 .\label{bclinstab}
\end{eqnarray}
These linear equations determine all $\hat{e}_i$ as a function of $\lambda$.
The fixed bias condition 
\begin{equation}
\sum_{i=1}^N  \hat{e}_i = 0, \label{delta-bias}
\end{equation}
then determines the possible eigenvalues $\lambda$.
For $\lambda\neq 0$ we now introduce the variable
$Y_i=\lambda\, \hat{e}_i/\hat{j}$ and the parameters
\begin{eqnarray}
b_i=I^* \frac{v'(E_i^*)}{v(E_i^*)}\, , \quad a_i=\frac{v(E_i^*)}{\nu}\, , 
\label{eqab}
\end{eqnarray}
in Eqs.~(\ref{Eqlinstab}) and (\ref{bclinstab}). Then we obtain:
\begin{eqnarray}
Y_i&=&\frac{\lambda +a_iY_{i-1}}{\lambda +a_i+b_i} \quad 
\mbox{or}\quad \left\{ 
\begin{array}{c}
\lambda +a_i+b_i=0\\ \mbox{and} \quad Y_{i-1}=-\lambda/a_i
\end{array} \right\} \label{eqy}\\
Y_1&=&\frac{\lambda}{\lambda+b_1}	
\end{eqnarray}

 Within the scope of this linear stability analysis, we find that
stationary states having all their fields $E_i$ in the 
positive differential mobility region are stable, which coincides with our 
physical intuition. This is formulated in the following lemma:

\bigskip
\noindent {\bf Lemma 3.}
{\em If $b_i\ge 0$ holds for all $i=1,\dots N$ the
real parts of all eigenvalues have to be negative, i.e., the state is stable.
Furthermore the current-voltage characteristic exhibits a positive slope
$dI^*/d\Phi$.
}
\bigskip

We prove this Lemma by contradiction. Let us assume that $Re(\lambda)\ge0$ holds
with $\lambda\neq 0$. As $b_1\ge0$ we directly find that $Re(Y_1)>0$ and $|Y_1-1|
\le 1$. In order to satisfy the voltage condition (\ref{delta-bias}) we conclude 
that there must be at least one $Y_j$ with $Re(Y_j)<0$. This implies directly
that also $|Y_j-1|>1$ must hold. But we find:
\begin{equation}
|Y_i-1|=\frac{|a_i(Y_{i-1}-1)-b_i|}{|\lambda +b_i+a_i|}<
\frac{|a_i||Y_{i-1}-1|+|b_i|}{|b_i+a_i|}
\label{escape}
\end{equation}
Given that $|Y_1-1|\le1$, the last equation implies that $|Y_i-1|\le 1$ for all 
$i$. Thus, the case $Re(\lambda)\ge 0$, $\lambda\neq 0$ is excluded. 

For $\lambda=0$  we obtain 
\begin{equation}
b_{1}\hat{e}_1 = \hat{j}\qquad \mbox{and} \qquad
(b_{i}+a_{i})\hat{e}_i = \hat{j}+a_{i}\hat{e}_{i-1}\, .\label{Eqlambda0}
\end{equation}
Therefore all
$\hat{e}_i$ have the same phase as $\hat{j}$ and the voltage condition
(\ref{delta-bias}) cannot be satisfied 
unless $\hat{e}_i = 0, \,\forall i$ and $\hat{j}=0$, which is the
trivial case. In conlusion $\lambda=0$ is not an eigenvalue. 
Furthermore $\lambda=0$
describes the infinitesimal change 
along the curve of stationary states. 
Eq. (\ref{Eqlambda0}) tells us, that $\hat{j}/\hat{e}_i\geq 0,\, \forall i$.
Identifying $dI^*=\hat{j}$ and $d\phi=\sum\hat{e}_i$
we obtain a positive slope of the current-voltage 
characteristic, i.e., $dI^*/d\phi\geq 0$.$\bullet$

\subsection{Stability for sufficiently large $\nu$}
Lemma 3 establishes that stationary field profiles with $v'(E^*_i)\geq 0$, 
$\forall i$ are linearly stable. These profiles include: (i) trivial ones 
where all the $E_i$ belong to the same branch of $v(E)$, and (ii) profiles 
where the negative differential mobility region is crossed in a single jump. 
This means that for a certain value $E_j^* \leq 1$ ($1\le j\le N-1$) 
there exists $E^*_{j+1}\geq E_m$ with $f(E^*_{j+1},I^*)=E_j^*$.
As $f(E,I^*)$ is increasing 
for $E\ge E_m$, the necessary and sufficient condition for the existence of
such a value $E^*_{j+1}$ is $f( E_m,I^*)\le E_j^*$. This condition yields
\begin{equation}
(E_m-E_j^*)v(E_m)\le \nu (I^*-v_m)
\end{equation}
This inequality is first fulfilled for $E_j^*=1$ and $I^*=1$,
which are the largest values of the respective quantities for the low
field domain. This gives:
\begin{equation}
\nu\ge\frac{(E_m-1)\, v_m}{1- v_m} \label{nustable}
\end{equation}
If this condition is fulfilled, domain states are possible
which cross the negative differential mobility region in a single jump
and must be stable therefore. 

Nevertheless there can be stable states even for smaller doping $\nu$,
as Lemma 3 only yields a sufficient and not a necessary condition
for stability. 
For our $v(E)$ curve inequality (\ref{nustable}) yields 
$\nu\ge 0.32$. 
Indeed we do not find any self-sustained 
oscillations for $\nu$ larger than the value $\nu\approx 0.27$, which is
somewhat smaller than our estimation. Checking other $v(E)$ 
curves we have always 
found oscillations up to a doping roughly $15-40\%$ lower than
determined by the bound~(\ref{nustable}).
Thus, the bound  is not only a sufficient condition but
also a reliable rough estimate for doping above which the oscillations
disappear.

Transforming  to physical units we obtain a surface charge density per well:
\begin{equation}
\tilde{N}_{D} \geq 
\frac{\tilde{v_m}\, \epsilon\, (\tilde{E_m} - \tilde{E_M})}
{(\tilde{v_M}-\tilde{v_m})\, q}\, ,
\end{equation}
which should be a reasonable approximation for the necessary doping density.
For the model equations from Prengel\cite{PRE94}  
we find $\tilde{N}_{D}\ge 2.4\times
10^{11}/cm^2$. Actually, oscillations are found in the model up to roughly 
$\tilde{N}_{D}\approx 10^{11}/cm^2$ for the regular 
superlattice and for
somewhat higher values for a slight amount of disorder \cite{WAC95b}.

If we regard domains with the high-field domain located at the injecting contact
the same arguments yield a bound 
\begin{equation}
\nu\ge\frac{(E_m-1)}{1-v_m} 
\end{equation}
for the existence of stable domains which is larger by a factor $1/v_m$.
For our $v(E)$ curve this corresponds to $\nu\ge 0.97$.
Indeed we have found stable domains with an depletion layer
for $\nu=1.0$ as depicted in Fig.~\ref{Figfield}. This indicates that this 
type of domain only appears for larger doping. This might explain that two 
different locations of the high-field domain
have been reported in the literature. In Ref.~\cite{HEL90} it is found to be
located at the injecting contact for a superlattice with 
$\tilde{N}_{D}=8.75\times 10^{11}$ cm$^{-2}$
while in Ref.~\cite{KWO95b} the high-field domain
is located at the receiving contact for a different superlattice with 
$\tilde{N}_{D}=1.5\times 10^{11}$ cm$^{-2}$.

\subsection{Stability for sufficiently small $\nu$}

Now we want to show that for sufficiently small doping, the (connected) branches 
of stationary solutions are stable. Then no self-sustained oscillatory branches
bifurcating from them can exist. In order to do this, we note that for very 
low voltages the stationary state is stable as indicated by Lemma 3: all field
values of this state are in the range $0<E^*_{i}<1$. We now increase the 
voltage and study whether an instability may occur by checking whether it is
possible to have $\lambda=i\omega$ with $\omega>0$ for some $\phi$.
(The case $\lambda=0$ yields the saddle-node bifurcation at the point
with $d\phi^*/dI^*=0$, which causes the switching to another branch
of the Z-shaped characteristic but typically does not generate
any oscillatory behavior.)
In the appendix we show that this is possible only if 
\begin{equation}
(N-1)\nu>\mbox{min}
\left\{ \frac{\pi\, (v_{l}-\nu c_1)}{4 c_1}\, ,\,
\frac{v_{l}\, c_1}{2 C\, (I^{*} - v_{l})c_2} \right\}
\label{EqNnu}
\end{equation}
holds, with
\begin{eqnarray}
v_{l}&=&\mbox{min}\left\{v_m\, ,\, (1+c)^{-1}\right\} \nonumber \\
c_1&=&I^*\mbox{max}_{E_l\le E\le E_h}|\partial\ln v(E)/\partial E|\nonumber \\
c_2&=&I^*\mbox{max}_{E_l\le E\le E_h}|\partial^{2}\ln v(E)/\partial E^{2}|
\nonumber \\
C &=& \frac{v_{l}}{(N-1)\nu c_1}\,\left[\exp\left(\frac{(N-1)\nu c_1}{v_{l} - 
\nu c_1}\right)- 1\right] \nonumber \, .
\end{eqnarray}
Here $E_l,E_h$ denote the minimal and maximal values of the field for the 
stationary field profile.
Note that for small $\nu$ we find $C\to v_l/(v_l-\nu c_1)$ and furthermore
the terms $\nu c_1$ become negligible. Then the right side no longer 
depends on $\nu$ and $N$ but only on the shape of $v(E)$ and the 
parameter $c$.

If $\nu$ is smaller and the inequality (\ref{EqNnu}) is violated,
no bifurcating oscillatory branches can issue forth 
from the steady state which is thereby stable.

The bound (\ref{EqNnu}) is far too small due to the rough estimations made 
during its derivation. Therefore the number itself should not be 
used for quantitative investigations. Nevertheless we now have shown that the 
stationary states are stable for low doping
and that in the limit of long superlattices the critical doping decreases as
$1/N$.

\subsection{Consequences for the continuum limit}
 With respect to  the continuum limit, $N\to\infty,\,\nu\to 0,\, L:=
N\nu < \infty$ we directly find that
there exists a minimal length $L_m$ such that the stationary
state is linearly stable if $L<L_m$. This lower bound is given by the
Eq.~(\ref{EqNnu}) with $\nu=0$, $C = 1$ and it
can be derived directly from the equations valid in the continuum limit, as we 
shall report elsewhere \cite{WAC96}. Eq.~(\ref{EqNnu}) and
similar bounds derived for other boundary conditions constitute an {\em 
explicit} form of the well-known $N_{3D}\tilde{L}$ criterion of the Gunn effect
\cite{SHA79}:
The dimensionless length $L$ (proportional to doping times the semiconductor
length \cite{HIG92,BON94b}) has to be larger than a certain number for the stationary 
solution to be unstable. 

Obviously the upper bound in the doping $\nu $ (for the absence of the
oscillatory regime) does not exist in the continuum limit ($\nu\to 0$, $N
\to\infty$, $L=N\nu$ fixed). The discreteness is essential for the field 
distribution to jump from the low-field region to the high field region 
without any fields exhibiting negative differential $v(E)$ in between, 
(which stabilizes the field distribution).
This explains that these stable stationary domains can not be found in
the usual Gunn diode: for large 
enough dc voltage bias, the Gunn diode may have a stable stationary 
solution with a large field near the anode.\cite{GUE71,SHA79}
However inside the diode there will be a region where the field takes values 
on the branch of negative differential velocity, which is different from 
what happens in the SL.

\section{Conclusions}

In this paper we have shown how the complex stationary current-voltage 
characteristic exhibiting domain branches is generated continuously as the
doping increases.
For low doping the characteristic follows the local $v(E)$ relation. If more 
charges are present,  the characteristic becomes Z-shaped. When the doping
is even larger, wiggles appear. For each doping the characteristic is connected, 
and the field profiles of all its different branches are monotone. The different 
disconnected branches observed experimentally correspond to the stable solution
branches of the full stationary current-voltage characteristic. It would be very
interesting to investigate whether it is possible to stabilize the unstable 
branches so that the full characteristic could be observed, as in the case of 
the double-barrier resonant-tunneling diode \cite{MAR94,WAC95a}.
Additionally, for large doping there exist isolated branches  on the 
full current-voltage characteristic having non-monotonic field profiles.

The stability analysis shows that the almost uniform field profile
is stable for low doping. The critical doping above which time-periodic
oscillations of the current appear is inversely proportional to the sample 
length for fixed superlattice parameters. This is the same situation as in the 
famous $N_D L$ criterion for the Gunn Diode. For yet larger doping the
time-periodic oscillations of the current disappear: there is an upper critical 
doping above which there appear stable stationary solutions with two electric
field domains (separated by an abrupt domain wall extending almost 
one period of the superlattice). 
Obviously, this is not possible for the conventional Gunn Diode
due to the lack of discretization.
It is important to mention that the upper critical doping needed to
stabilize stationary domain structures is higher for profiles having
depletion layers instead of accumulation layers between the
different domains.

This paper gives (for the first time) analytical bounds of the interval of
dimensionless doping (proportional to 2D doping over
field at resonant tunneling peaks) on which self-sustained current
oscillations may exist. To ascertain the influence of the SL parameters 
and the voltage bias on the current oscillations themselves, a detailed
analysis of the dynamics of the model is necessary. This analysis together
with experimental verification and predictions on the possibility of
tuning the oscillation frequency with applied bias will be presented
in a companion paper elsewhere. \cite{KAS96}

\acknowledgements
We thank J. Gal\'an, H. T. Grahn, J. Kastrup, M. Patra, F. Prengel, 
G. Schwarz, E. Sch{\"o}ll and S.\ Venakides for fruitful discussions and 
collaboration on related topics. We thank E. Doedel for sending us his program 
of numerical continuation AUTO. One of us (AW) acknowledges financial 
support from the Deutsche Forschungsgemeinschaft (DFG).
This work has been supported by the DGICYT grant PB94-0375, and by the EU 
Human Capital and Mobility Programme contract ERBCHRXCT930413.

\appendix

\section{Proof of stability for sufficiently small $\nu$}

Here we prove that $\lambda=i\omega$ with $\omega>0$ can be an eigenvalue
of the linearized system (\ref{Eqlinstab}) only if $\nu$ is sufficiently large.

In order to do this, we assume that a given stationary field profile 
$\{E^c_i\}$ exhibits $\lambda=i\omega$ with $\omega>0$ and
derive several necessary  conditions for this.
Restricting ourselves to increasing field 
profiles, $E^c_1$ must either be located on the first
or second branch of the $v(E)$ curve, as otherwise
the second branch is not reached which is a necessary condition
for the instability according to Lemma 3.
Let us now determine the smallest value $E_l$ 
and the largest value $E_h$
the  stationary field profile $\{E^c_i\}$ may take. $E_l$
is given by the value of $E^c_1$ on the first branch 
for which the current takes on its minimal value, $I^*=1$, considering that
the field must eventually take values on the second branch of $v(E)$. 
Eq.~(\ref{Eqinitials}) yields $v(E_l)=1/(1+c)$ which determines 
the field  $E_l\le 1$.
$E_h$ is given by the largest value that $E^c_N$ 
can take on the third branch of 
$v(E)$. Noticing that $I^*\le 1+c$ in Eq.~(\ref{Eqinitials}), we can adopt 
$E_h$ as the solution of $v(E_h)=1+c$ from the third branch.
Thus, $E_l$ and $E_h$ depend only on the $v(E)$ curve and the parameter $c$
but not on the field profile \{$E^c_i$\}.
For sake of convenience we introduce the following quantities:
\begin{eqnarray}
v_l&: =&\mbox{min}\left\{v_m,\frac{1}{1+c}\right\}\\
c_1&: =&I^{*}\, \mbox{max}_{E_l\le E\le E_h}
\left| \frac{\partial\ln v(E)}{\partial E}\right|\\
c_2&: =&I^{*}\, \mbox{max}_{E_l\le E\le E_h}
\left| \frac{\partial^2 \ln v(E)}{\partial E^2}\right|
\end{eqnarray}
Then  we have
\begin{equation}
a_i\ge \frac{v_l}{\nu} \quad \mbox{and} \quad
|b_{i}|\le c_1\quad \forall i
\label{Eqbmax}
\end{equation}
In the following we will assume that $\nu$ is so small that the 
function $f(E,I^*)$ ($I^*$
fixed) is always invertible and furthermore 
\begin{equation} 
a_i+b_i > v_l/\nu - c_1 > 0\, ,\quad\forall i
\label{restriccion}
\end{equation}
holds.

For $\lambda = i\omega$, $\omega>0$, 
we have Re$Y_1>0$, and $|Y_1 -1|<1$. As previously 
explained in the proof of Lemma 3, there must be a $Y_j$ such that 
Re$Y_i\ge 0$, $i= 1, \ldots, j-1$, and Re$Y_j<0$ in order to fulfil the 
voltage condition. We are going to prove the following result:

\noindent {\bf Lemma 4.}
{\em Let $j>1$ be the index that satisfies Re$Y_i\ge 0$, $i = 1, \ldots,j-1$, and 
Re$Y_j<0$. \\
(a) If $\omega \le c_1$, 
we have 
\begin{eqnarray}
(j-1)\, \nu > \frac{\pi v_{l}}{4 A c_1}\label{ineqA}
\end{eqnarray}
where A is the maximum of the expressions
\begin{eqnarray}
A_k:= \frac{v_{l}}{(j-k+1)\nu}\,\sum_{i=k}^{j} \frac{1}{a_{i}+b_{i}}
\, ,\label{A}
\end{eqnarray}
for $k=2,\ldots j$.\\
(b) If $\omega >c_1$, we have
\begin{eqnarray}
(j-1)\,\nu > 
\frac{ v_{l}c_1}
{2B \, (I^{*} - v_{l})\, c_2} \, ,
\label{ineqB}
\end{eqnarray}
where 
\begin{eqnarray}
B&:=& \frac{\mu\,(\mu^{j-1}-1)}{(\mu-1)\, (j-1)}\, ,\quad\mbox{with} \label{B}\\
\mu&:=& \mbox{max}_{E_l\le E_i\le E_h}\left\{ \frac{a_{i}}{|a_{i} + b_{i} 
+ i c_1|} \right\}\, . \label{mu}
\end{eqnarray}
}
\bigskip

\noindent {\em Proof}:

\underline{{\bf (a)} Let $\omega \le c_1$:}\\
In order to prove (\ref{ineqA}), we consider how the 
argument $\phi_i$ of the complex quantity $Y_i$ is varying with $i$.
\begin{equation}
\frac{Y_{i-1}}{Y_{i}}=\frac{|Y_{i-1}|}{|Y_{i}|}e^{i(\phi_{i-1}-\phi_{i})}
=\frac{1}{a_i}\left(i\omega(1-Y_{i}^{-1}) +b_i+a_i \right)
\end{equation}
Therefore we get:
\begin{equation}
\phi_{i-1}-\phi_{i}=\arctan \left(\frac{\omega-\omega Re(Y_{i}^{-1})}
{b_i+a_i+\omega Im(Y_{i}^{-1})} \right)
\end{equation}
Furthermore we have:
\begin{equation}
\phi_{1}=\arctan\left(\frac{b_1}{\omega}\right) \label{Eqphi1}
\end{equation}
Straightforward calculations starting from eq.~(\ref{eqy}) yield:
\begin{eqnarray} 
Re(Y_{i})&=&\frac{\omega^2+a_i\omega Im(Y_{i-1}) +a_i(b_i+a_i)Re(Y_{i-1})}
{\omega^2+ (b_i+a_i)^2} \nonumber\\
Re(Y_{i})&=&\frac{\omega}{b_i+a_i}Im(Y_{i})+
\frac{a_i}{b_i+a_i}Re(Y_{i-1}) \label{equalY_i}
\end{eqnarray}
By definition of the index $j$ ($j\le N$), Re$Y_{j}<0$ and Re$Y_{j-1}\ge 0$.
Then these equations indicate that Im$Y_{j}<0$ and Im$Y_{j-1}<0$. 
Thus the 
transition Re$Y_{j-1}\ge 0\, \to$ Re$Y_{j}<0$ occurs across the angle
$\phi=-\pi /2$ as we have 
$-\pi /2 \le \phi_{j-1}<0$ and $-\pi<\phi_{j}<-\pi /2$.

We introduce the index $j'$ which is defined by the relations
$-\pi/4\le \phi_{j'-1}$ and $\phi_i< -\pi/4$ for $i=j',\ldots,j$.
Obviously, we have Im$Y_{i}<0$ and therefore 
$b_i+a_i-\omega \frac{Im(Y_{i})}{|Y_{i}|^{2}}>0$  
for $i=j',j'+1,\ldots,j$. 
Using (\ref{equalY_i}) and Re$Y_{i-1}\ge 0$
(for all $i\leq j$), we obtain Re$Y_i \ge \omega$ Im$Y_i/(a_i + b_i)$. 
This  yields for $i=j',\ldots,j$:
\begin{eqnarray}
\phi_{i-1}-\phi_{i}&=&
\arctan \left(\frac{\omega-\omega \frac{Re(Y_{i})}{|Y_{i}|^{2}}}
{b_i+a_i-\omega \frac{Im(Y_{i})}{|Y_{i}|^{2}}} \right)\nonumber\\
&\le& \arctan \left(\frac{\omega-\frac{\omega^{2} Im(Y_{i})}
{(b_i+a_i)|Y_{i}|^{2}}}
{b_i+a_i-\omega \frac{Im(Y_{i})}{|Y_{i}|^{2}}} \right)\nonumber\\
&=&\arctan\left( \frac{\omega}{a_i+b_i}\right)
<\frac{\omega}{a_i+b_i}\label{Eqdiff}
\end{eqnarray}
 Now we have to distinguish two different cases:

\noindent i) $j'\ge2$: 
By summing the inequality (\ref{Eqdiff}) from $i=j'$ to $i=j$ and then taking 
into account the definitions of $j$ and $j'$, we find:
\begin{eqnarray} 
-\frac{\pi}{4}+\frac{\pi}{2}&<&\phi_{j'-1}-\phi_{j}
<\sum_{i=j'}^{j} \frac{\omega}{a_i+b_i}\nonumber \\
&=& \frac{\omega\, (j-j'+1)\,\nu\, A_{j'}}{v_{l}} \, ,
\label{sumab}
\end{eqnarray}
where definition (\ref{A}) has been used. The property $\omega \le c_1$
then implies
\begin{equation} 
j-1\ge j-j'+1 > \frac{\pi v_{l}}{4A_{j'}\nu c_1}
\label{eqA1}\, .
\end{equation}
 
\noindent ii) $j'=1$: This means that $\phi_1<-\pi/4$ and according to
to Eq.~(\ref{Eqphi1}), $\omega<-b_1=|b_1|$. 
Now we sum the inequality (\ref{Eqdiff}) from $i=2$ to $i=j$ and then use
the expression $\arctan x > \pi x/4$ for $0<x<1$, thereby obtaining 
\begin{eqnarray} 
\frac{\pi\omega}{-4b_1}&<&\arctan\left(\frac{\omega}{-b_1}\right) =
\arctan\left(\frac{b_1}{\omega}\right)+\frac{\pi}{2}\nonumber \\
&<&\phi_1-\phi_j
<\frac{\omega\, \nu \, (j-1)\, A_2}{v_{l}}\, .
\end{eqnarray}
Therefore we find with Eq. (\ref{Eqbmax}):
\begin{equation} 
j-1 > \frac{\pi v_{l}}{4\, \nu \, A_2\, |b_1|}
\ge  \frac{\pi v_{l}}{4\, A_2\, \nu c_1}\label{eqA2}\, .
\end{equation}

Putting together (\ref{eqA1}) and (\ref{eqA2}) 
we obtain the inequality  (\ref{ineqA}). 

\underline{{\bf (b)} Let now $\omega>c_1$:} \\
To prove the inequality (\ref{ineqB}), 
we shall define the auxiliary functions
\begin{equation} 
Z_i = Y_i - \frac{\lambda}{\lambda +b_{i}}\label{defZ}\, .
\end{equation}
These functions solve the following discrete equation
\begin{equation} 
(\lambda + a_i + b_i)\, Z_i - a_i \, Z_{i-1} =  \lambda a_i \,\left(\frac{1}
{\lambda + b_{i-1}} - \frac{1}{\lambda + b_{i}}\right) \, ,\label{eqZ} 
\end{equation}
with the boundary condition $Z_1 = 0$. The solution of this problem is
\begin{equation} 
Z_n = \sum_{k=2}^{n} \frac{\lambda\,(b_{k} - b_{k-1})}{(\lambda + b_{k-1})\,
(\lambda +b_{k})} \, \prod_{i=k}^{n}\frac{a_{i}}{\lambda + b_{i} + a_{i}}
\label{solZ}\, .
\end{equation}
As all $b_i$ are real quantities, we have $|i\omega +b_i|>\omega$ 
and obtain the following inequality for 
$|Z_j|$ by using the preceding formula with 
$\lambda = i\omega$,
$\omega > 0$:
\begin{equation} 
|Z_j| <  \frac{1}{\omega}\, \sum_{k=2}^{j} |b_{k} - b_{k-1}|\,
\prod_{i=k}^{j}\frac{a_{i}}{|b_{i} + a_{i} + i\omega |}
\label{ineqZ1}\, .
\end{equation}
Now we have $|Z_j|> -$ Re$Z_j = -$ Re$Y_j + \omega^2/(\omega^2 + b_j^2) >
\omega^2/(\omega^2 + b_j^2)$, where Re$Y_j<0$ and the definition of $Z_n$
have been used. This inequality together with (\ref{ineqZ1}) yield
\begin{equation} 
\frac{\omega^{3}}{\omega^{2} + b_{j}^{2}} <  \sum_{k=2}^{j} |b_{k} - b_{k-1}|\,
\prod_{i=k}^{j}\frac{a_{i}}{|b_{i} + a_{i} + i\omega |}
\label{ineqZ2}\, .
\end{equation}

We now estimate the right side of (\ref{ineqZ2}). The definition (\ref{eqab}) 
of $b_i$ and the mean value theorem yield 
\begin{eqnarray}
|b_k - b_{k-1}| < c_2\, |E_k^c - E_{k-1}^c| .
\label{mvt}
\end{eqnarray}
Equation (\ref{Eqfdef})
for the stationary state now yields $0 < E_k^c - E_{k-1}^c = (I^*/a_k - \nu)$,
so that $0< E_k^c - E_{k-1}^c < \nu (I^*/v_l - 1)$.
Thus we can write:
\begin{eqnarray}
|b_k - b_{k-1}| <\nu \left( \frac{I^{*}}{v_{l}} - 1\right) \, c_2\, . 
\label{ineqb_k}
\end{eqnarray}
On the other hand, as we are considering the case $\omega > c_1$, we find 
that
\begin{eqnarray}
\frac{a_{i}}{|b_{i} + a_{i} + i\omega |} < \frac{a_{i}}{|b_{i} + 
a_{i} + i c_1 |}\leq \mu ,
\label{ineq_mu}
\end{eqnarray}
according to (\ref{mu}). Inserting (\ref{ineqb_k}) and (\ref{ineq_mu}) into 
(\ref{ineqZ2}), we obtain
\begin{eqnarray} 
\frac{\omega^{3}}{\omega^{2} + c_1^{2}} 
&<& \nu \, c_2\, \left( 
\frac{I^{*}}{v_{l}} - 1\right)\, \left(\sum_{k=2}^{j}\mu^{j-k+1}\right)\, 
\nonumber\\
&=& \nu \, c_2 \, \left( \frac{I^{*}}{v_{l}} - 1\right)\, \mu\,
\frac{\mu^{j-1}-1}{\mu -1}\, . \label{ineqZ3}
\end{eqnarray}
Since $\omega > c_1$, $\omega^{3}/(\omega^{2} +c_1^{2}) >
c_1/2$. Inserting this into (\ref{ineqZ3}), we obtain (\ref{ineqB}).
Therefore Lemma 4 is proved.$\bullet$

Lemma 4 yields necessary conditions for the instability of a given stationary
field profile $\{E_i^{c}\}$ corresponding to a fixed bias. We would like 
to obtain a general condition on $\nu$, which should only depend on the
$v(E)$ curve and the parameter $c$, but not on the specific stationary 
field profile. This can be achieved by the following considerations:
\begin{eqnarray}
A_k&=& \frac{v_{l}}{(j-k+1)\nu}\,\sum_{i=k}^{j} \frac{1}{a_{i}+b_{i}}\le 
\frac{v_{l}}{v_{l}-\nu c_1}\, .
\end{eqnarray}
Therefore we have $A<v_{l}/(v_{l}-\nu c_1)$ from Eq.~(\ref{A}), which 
inserted into the inequality (\ref{ineqA}) gives 
\begin{eqnarray}
(j-1) \nu > \pi \frac{v_l -  \nu c_1}{4c_1}\,  .
\label{uno}
\end{eqnarray} 
>From the definition (\ref{mu}) we obtain
\begin{eqnarray}
\mu \le \mbox{max}_{E_l\le E_i\le E_h}\left\{
\frac{a_{i}}{a_{i} - c_1}\right\}
\le 1 +  \frac{\nu c_1}{v_{l} - \nu c_1}\, .
\end{eqnarray}
This yields
\begin{eqnarray}
B &\le& \frac{v_{l}}{(j-1) \nu c_1}\,\left[\left( 1 +  \frac{\nu c_1}{v_{l} - 
\nu c_1}\right)^{j-1} - 1\right]\nonumber \\
&\le& \frac{v_{l}}{(j-1) \nu c_1}\,\left[\exp\left(\frac{(j-1)\nu c_1}{v_{l} - 
\nu c_1}\right) - 1\right]\, =: C\, ,
\end{eqnarray}
to be inserted in (\ref{ineqB}). The result is
\begin{eqnarray}
(j-1)\,\nu > 
\frac{ v_{l}c_1}
{2C \, (I^{*} - v_{l})\, c_2} \, ,
\label{dos}
\end{eqnarray}

We now use the obvious inequality $N\ge j$ in (\ref{uno}) and (\ref{dos})
thereby obtaining the condition (\ref{EqNnu}) as a necessary condition for
oscillatory instability of the steady state.

\begin{figure}
\caption{Current-voltage characteristics for $c=0.01$ and $N=20$ 
and different values of $\nu$. 
The full line denotes the states where the electric field $E_i$
is strictly monotone increasing in $i$. For $\nu=1.0$ there appear 
additional branches with non-monotonic field profiles $E_i$. They are
isolated from the stationary branches having increasing 
field profiles.
We have shown one such branch, which has also been blown up for the sake of 
clarity. The dotted line is the $v(E)$ curve used throughout this paper.}
\label{Figchars}
\end{figure}

\begin{figure}
\caption{$f(E,I)$ and a trajectory $E_i$ indicating decreasing $i$ for
various doping densities  and currents.
(a) $\nu=0.15$, $I=0.8$. (b) $\nu=1.0$, $I=0.8$. (c) $\nu=1.0$, $I=1.008$.
(d) $\nu=1.0$, $I=0.599=I_c$.}
\label{FigfofE}
\end{figure}

\begin{figure}
\caption{Different stationary electric field profiles for $c=0.01$, $\nu=1.0$,
$\phi=1.2$, and $N=20$. The crosses mark a state of the connected branch
from Fig.~1($\nu=1.0$) while the circles mark a 
state belonging to the isolated branch.}
\label{Figfield}
\end{figure}

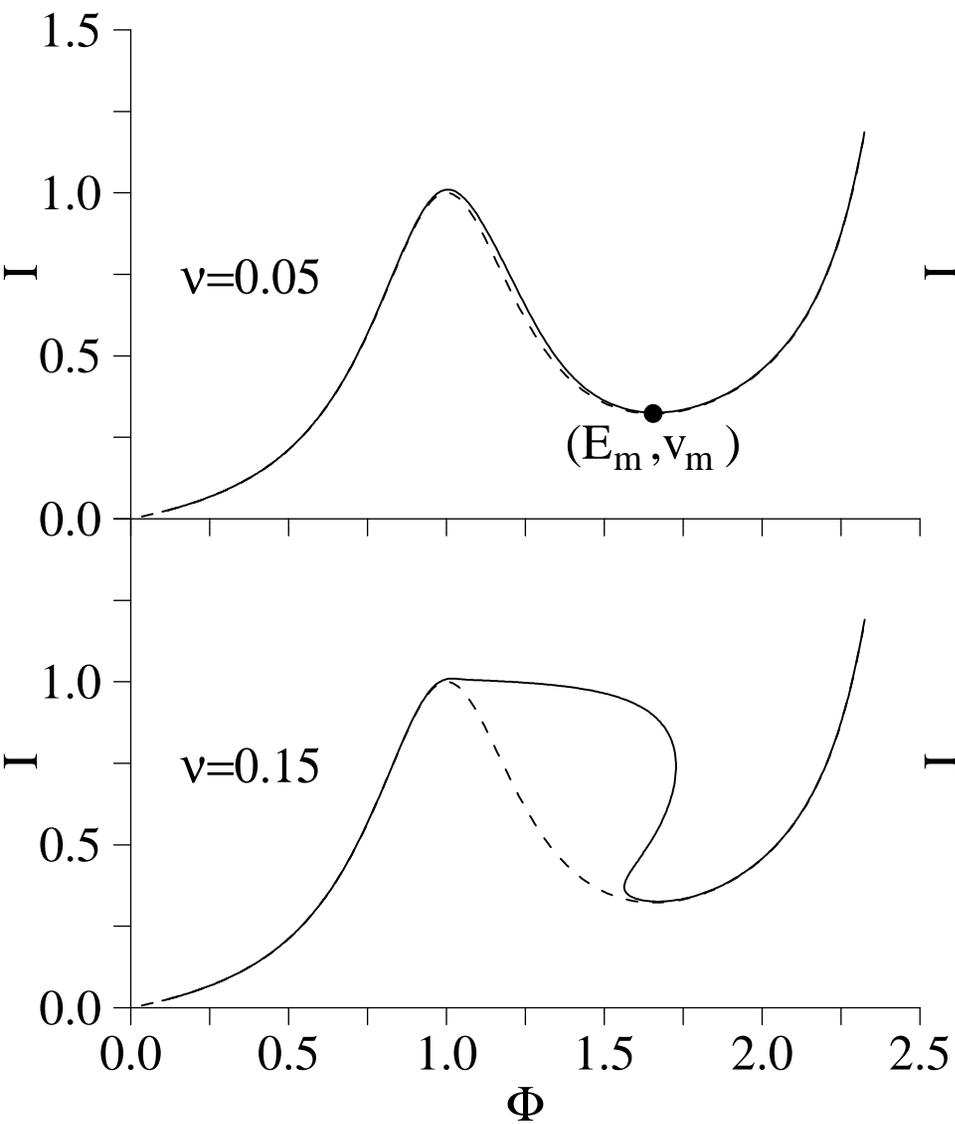
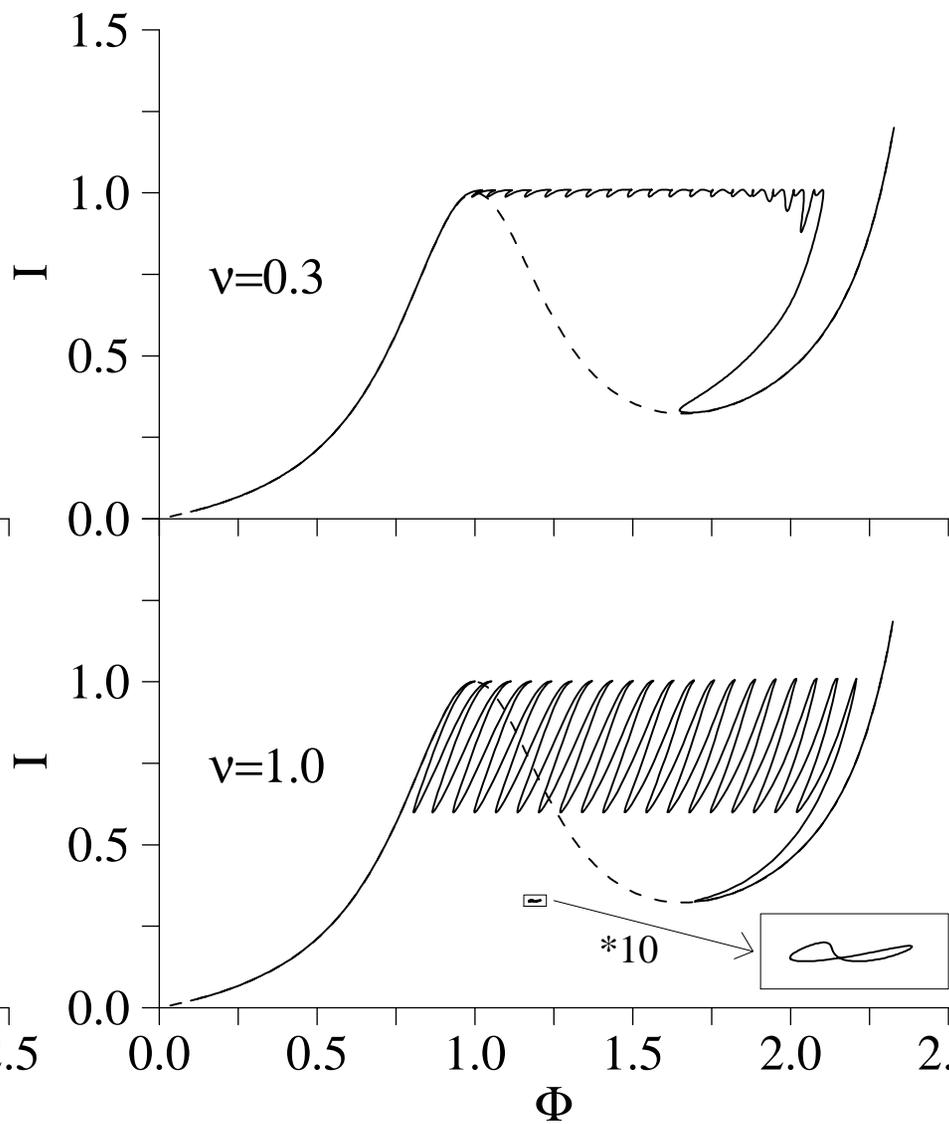
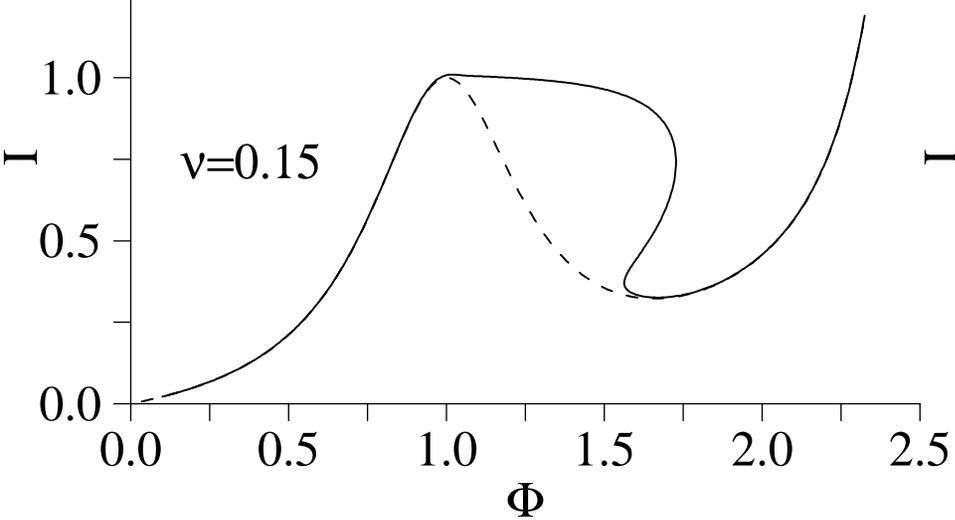
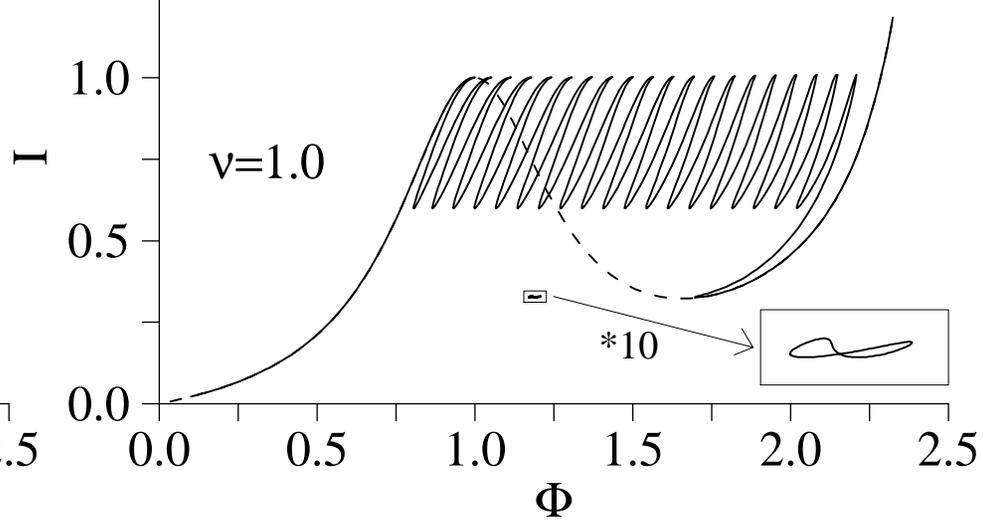
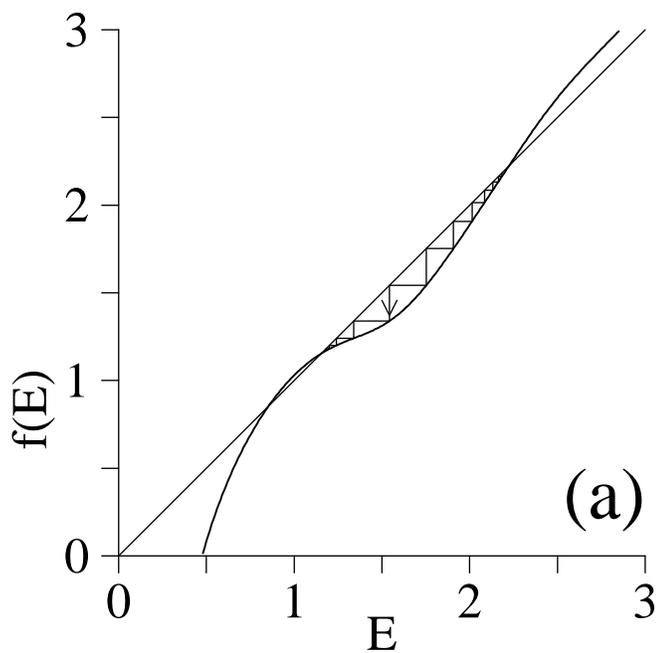
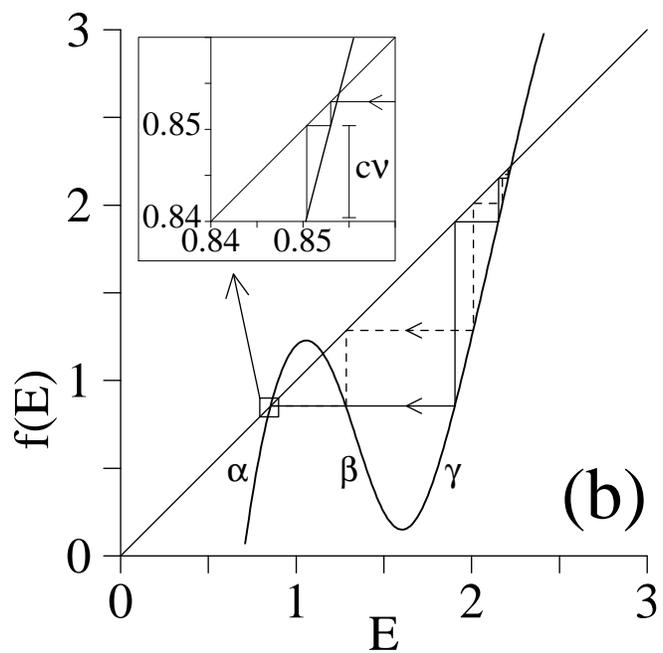
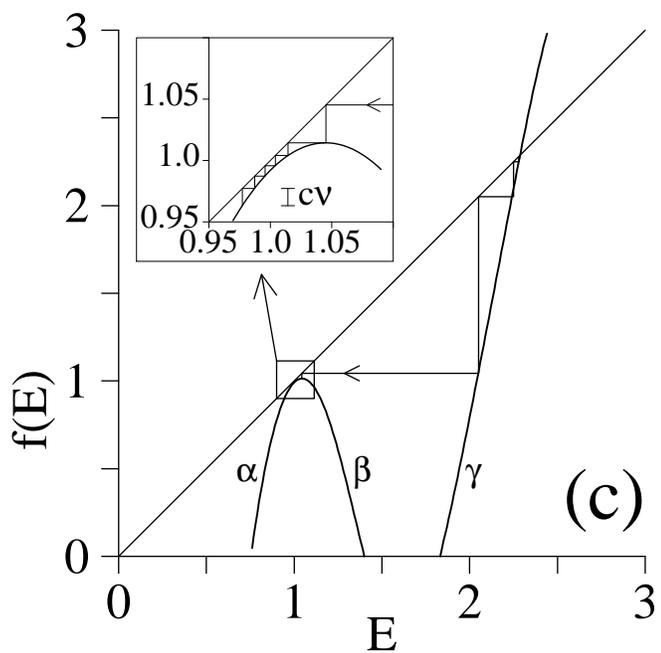
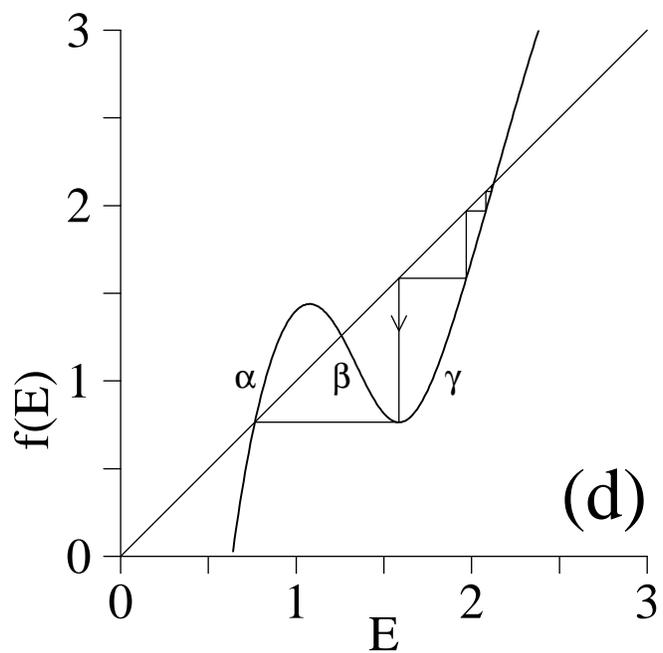
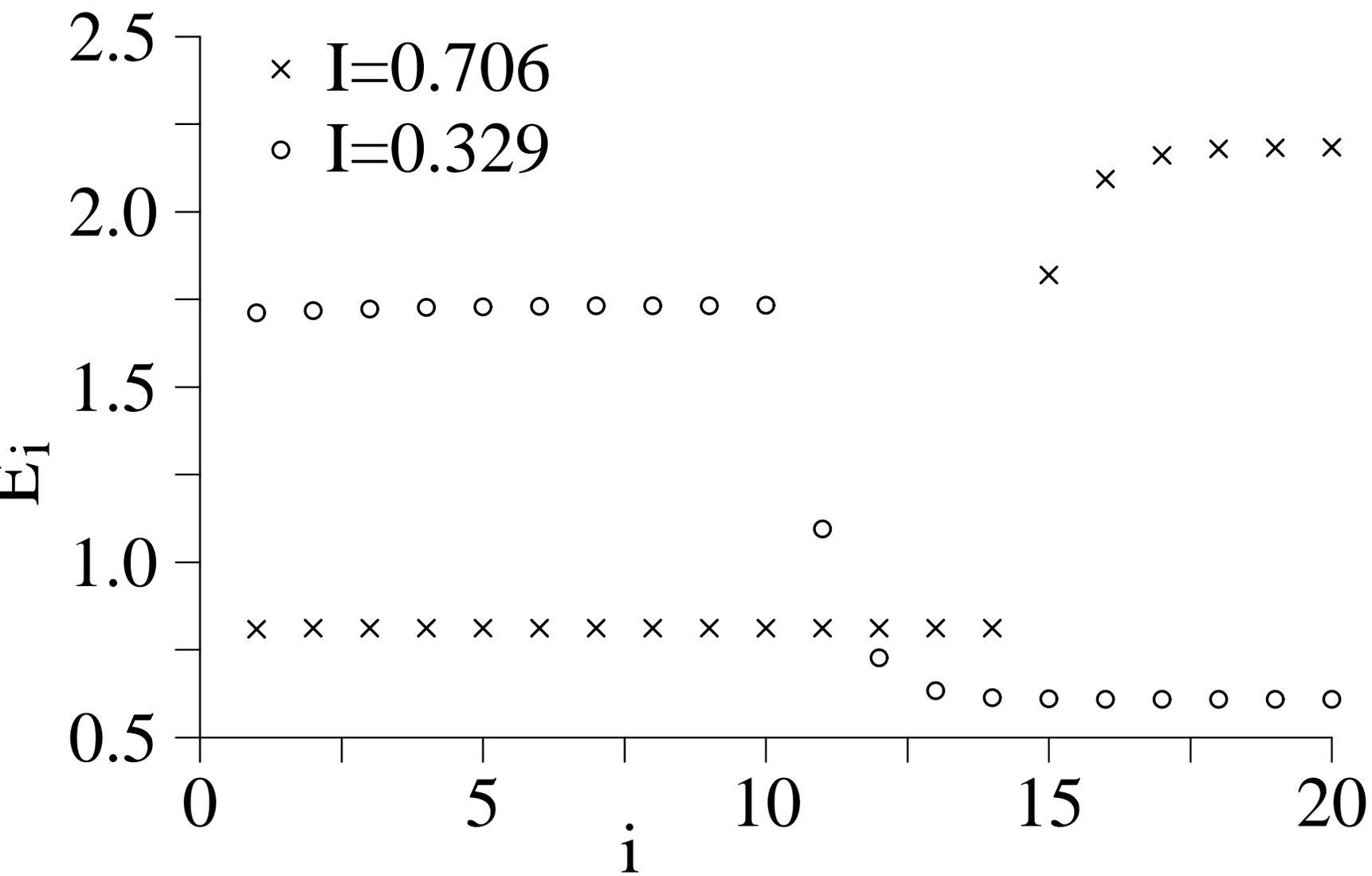
\end{document}